\documentclass{aa}
\usepackage{times}
\usepackage{astron}
\usepackage{psfig}
\begin{document}
\newcommand{\hb}{H$\beta$}
\newcommand{\hg}{H$\gamma$}
\newcommand{\mgt}{Mg$_2$}
\newcommand{\mgb}{Mg$_{\rm b}$}
\newcommand{\fef}{Fe\,5}
\newcommand{\fes}{Fe\,6}
\newcommand{\magn}{mag}

   \thesaurus{3(11.03.2; 11.05.1; 11.06.2)} 
   \title{M32--like galaxies: still very rare}

   \subtitle{M32 analogues do not exist in the Leo group}

   \author{B. L. Ziegler
                \thanks{Visiting astronomer of the German--Spanish
Astronomical Center, Calar Alto, operated by the Max--Planck--Institut f\"ur
Astronomie, Heidelberg, jointly with the Spanish National Commission for
Astronomy.}
   \and R. Bender
          }

   \offprints{B. L. Ziegler}

   \institute{Universit\"ats--Sternwarte, 
   Scheinerstra\ss e 1, 81679~M\"unchen, Germany\\
              email: ziegler@usm.uni-muenchen.de
             }

   \date{Received August, 1997; accepted October, 1997}

   \maketitle

   \begin{abstract}

New observations of galaxies in the direction of the Leo group, which were
classified as being candidates for M32--like dwarf galaxies by Ferguson \&
Sandage \cite*{FS90}, show that these galaxies are normal elliptical
galaxies at redshifts between $z=0.02$ and $z=0.06$. This result adds
further evidence that the faint end of the luminosity function of elliptical
galaxies falls off very steep and that objects like M32 are extremely rare.

      \keywords{galaxies: compact -- galaxies: elliptical and lenticular, cD
                -- galaxies: fundamental parameters}

   \end{abstract}

%________________________________________________________________

\section{Introduction}

M\,32 is a galaxy with very unusual characteristics. Compared to dwarf
galaxies having similar absolute brightness ($M_B$) its central surface
brightness ($\mu_o$) is 4 orders of magnitudes higher and its core radius
($r_c$) 3 orders of magnitude smaller \cite{Korme85}. Within the two planes
`$\mu_o \mbox{vs.} M_B$' and `$r_c \mbox{vs.} M_B$', M\,32 lies
approximately on the extension of the sequence defined by elliptical
galaxies and is clearly distinct from the low surface brightness dwarf
galaxies following a different sequence (cf. with Fig.\,\ref{sbm}).
Therefore, M\,32 is often denoted as a dwarf elliptical galaxy and the other
dwarf galaxies as dwarf spheroidals (dSph) \cite{Korme87,Djorg92}, whereas
other authors use the term compact elliptical for M\,32 and dwarf elliptical
for the low surface brightness objects only \cite{SB84,BBF93}). So far, only
very few other galaxies with properties similar to M\,32 have been found
\cite{SB84,NP87,David91,KB94}. But none of these galaxies has such extreme
properties as M\,32.

%Sandage \& Binggeli 

Ferguson \& Sandage \cite*{FS90} classified a number of galaxies in nearby
groups as possible M32-type objects due to their compact appearance on
photographic plates (similar to NGC 4486B in Sandage \& Binggeli
\cite*{SB84} though NGC\,4486B is still 2 \magn\ brighter than M\,32). We
obtained spectroscopy of these candidate galaxies in the Leo group during a
longer observing run within our project to study the evolution of elliptical
galaxies \cite{BZB96,ZB97}. The galaxies were drawn from Table\,II of
Ferguson \& Sandage \cite*{FS90} and are designated in the following as {\sf
Leo\,\#}, where {\sf \#} corresponds to the number assigned in that table.

In the following sections we briefly describe the observations and their
analysis, present the results and draw our conclusions.

%________________________________________________________________

\section{Observations and data analysis}

Spectroscopic observations were made at the 3.5m telescope on Calar Alto
using the Boller\&\-Chivens TWIN spectrograph. The spectra cover the
wavelength range $\lambda\lambda=4500-6100$\,\AA\ and have an instrumental
resolution of $\sigma_{\rm i}=105\mbox{km s}^{-1}$. The exposure time was
1\,h. The CCD images were reduced in the usual way and the spectra were
extracted and integrated with an algorithm introduced by Horne
\cite*{Horne86}. Absorption line indices of \hb, \mgt, \mgb, \fef\
($\lambda_0\approx5270$\,\AA) and \fes\ ($\lambda_0\approx5335$\,\AA) were
measured in the one-dimensional spectra following Faber et al.
\cite*{FFBG85}. They were corrected for velocity dispersion broadening and
normalized to the Lick system by means of observed reference stars. Velocity
dispersions were derived using the FCQ-method by Bender \cite*{Bende90a}.
Details of the observational set-up, the reduction procedure and the
derivation of the data can be found in Ziegler \& Bender \cite*{ZB97}.

\begin{table*}[ht]
\caption[]{Spectroscopic parameters}
\label{spec}
\begin{flushleft}
\begin{tabular}{lrrllllllll}
\noalign{\smallskip}
\hline
\noalign{\smallskip}
galaxy & $v_r$ & $\sigma$ & \mgb & $\Delta$\mgb & \fef & $\Delta$\fef & \fes
& $\Delta$\fes & \hb & $\Delta$\hb \\
 & (km\,s$^{-1}$) & (km\,s$^{-1}$) & (\AA) & (\AA) & (\AA) & (\AA) & (\AA) & 
 (\AA) & (\AA) & (\AA) \\
\noalign{\smallskip}
\hline
\noalign{\smallskip}
  Leo12 &  7299 & $\le$81 & 1.27 & 0.14 & 1.22 & 0.15 & 1.58 & 0.18 & 1.22 &
0.10\\
  Leo16 & 16494 & 385     & 4.82 & 0.08 & 3.00 & 0.09 & 1.92 & 0.10 & 1.81 &
0.06\\
  Leo18 & 14357 & 134     & 3.51 & 0.12 & 2.19 & 0.14 & 1.62 & 0.16 & 1.95 &
0.10\\
  Leo44 & 17992 & 259     & 2.61 & 0.10 & 2.86 & 0.11 & 2.42 & 0.12 & 1.73 &
0.07\\
  Leo51 &  6518 & 154     & 3.43 & 0.12 & 2.72 & 0.14 & 1.79 & 0.16 & 1.87 &
0.10\\
    M32 & --220 &  78     & 2.40 & 0.03 & 2.84 & 0.03 & 2.49 & 0.03 & 2.11 &
0.03\\
NGC1600 &  4973 & 382     & 5.25 & 0.03 & 3.04 & 0.04 & 3.19 & 0.04 & 1.50 &
0.03\\
NGC2300 &  1954 & 289     & 5.07 & 0.03 & 2.97 & 0.03 & 2.84 & 0.04 & 1.53 &
0.02\\
\noalign{\smallskip}
\hline
\end{tabular}
\end{flushleft}
\end{table*}

To get an estimate of the apparent size, surface brightness and total
brightness, the {\sf Leo} galaxies were imaged with a CCD camera at the
80\,cm telescope of the Wendelstein Observatory. A filter corresponding to
the Cousins $R$ band was used. A calibration accurate to about 0.1 \magn\
was achieved by the observation of the standard star cluster NGC\,4147
\cite{CABBHMS85}, which is located in the vicinity of the {\sf Leo} group.
%but no correction for air mass was made.
The effective radius ($R_e$), effective surface brightness at $R_e$ (SB$_e$)
and total apparent magnitude ($M_t$) were derived with the isophote fitting
technique introduced by Bender \& M\"ollenhoff \cite*{BM87} and using an
$r^{1/n}$ growth curve.

%________________________________________________________________

\section{Results and conclusions}

In addition to the program galaxies {\sf Leo\,12}, {\sf Leo\,16}, {\sf
Leo\,18}, {\sf Leo\,44} and {\sf Leo\,51}, spectroscopy was also obtained of
the giant elliptical galaxies NGC\,1600 and NGC\,2300 as well as of M\,32
itself for reference. Table\,\ref{spec} contains the spectroscopic data: the
heliocentric radial velocity ($v_r$), the velocity dispersion ($\sigma$) and
the absorption indices corrected for velocity dispersion broadening. The
measurement errors of $v_r$ and $\sigma$ are less than 10\,km\,s$^{-1}$.
Our instrumental resolution is too low to get a correct value of $\sigma$
for M\,32 and that's why we use the measurement of Davies et al.
\cite*{DBDFLTW87}. These authors used an aperture size which is effectively
the same as ours. 
%The spectra of the {\sf Leo} galaxies can be found in the appendix.???

For one M32 candidate -- {\sf Leo\,38} -- we did not succeed in getting a
spectrum but we obtained an image and it is included in Table\,\ref{phot}.
There, the effective radius ($R_e$), the effective surface brightness at
$R_e$ (SB$_e$) and the total apparent magnitude ($M_t$) in the $R$ band are
listed. To convert this magnitude to the $B$ band, we use a $(B-R)$ color
index of 1.8 which is typical for an elliptical galaxy at $z=0.04$
\cite{BC93}. We estimate the error in magnitude to be roughly 0.1\,\magn\
and in effective radius 0\,\farcs5. In Table\,\ref{phot}, we also indicate
the appearance of a substantial exponential component in the surface
brightness profile.

\begin{table}[h]
\caption[]{Photometric parameters}
\label{phot}
\begin{flushleft}
\begin{tabular}{lllll}
\noalign{\smallskip}
\hline
\noalign{\smallskip}
 galaxy & $R_e$ & SB$_e$ & $M_t(R)$ & profile\\
 & (\arcsec) & (\magn$/\sq\arcsec$) & (\magn) & \\
\noalign{\smallskip}
\hline
\noalign{\smallskip}
 Leo12 & 5.1 & 21.7 & 15.0 & disk\\
 Leo16 & 8.2 & 21.5 & 13.6 & \\
 Leo18 & 2.5 & 20.3 & 15.4 & \\
 Leo38 & 4.0 & 22.3 & 16.5 & disk\\
 Leo44 & 7.1 & 21.8 & 14.1 & \\
 Leo51 & 4.5 & 20.3 & 14.0 & disk\\
\noalign{\smallskip}
\hline
\end{tabular}
\end{flushleft}
\end{table}

\begin{figure*}
%\begin{center} \parbox{18cm}{ 
\psfig{figure=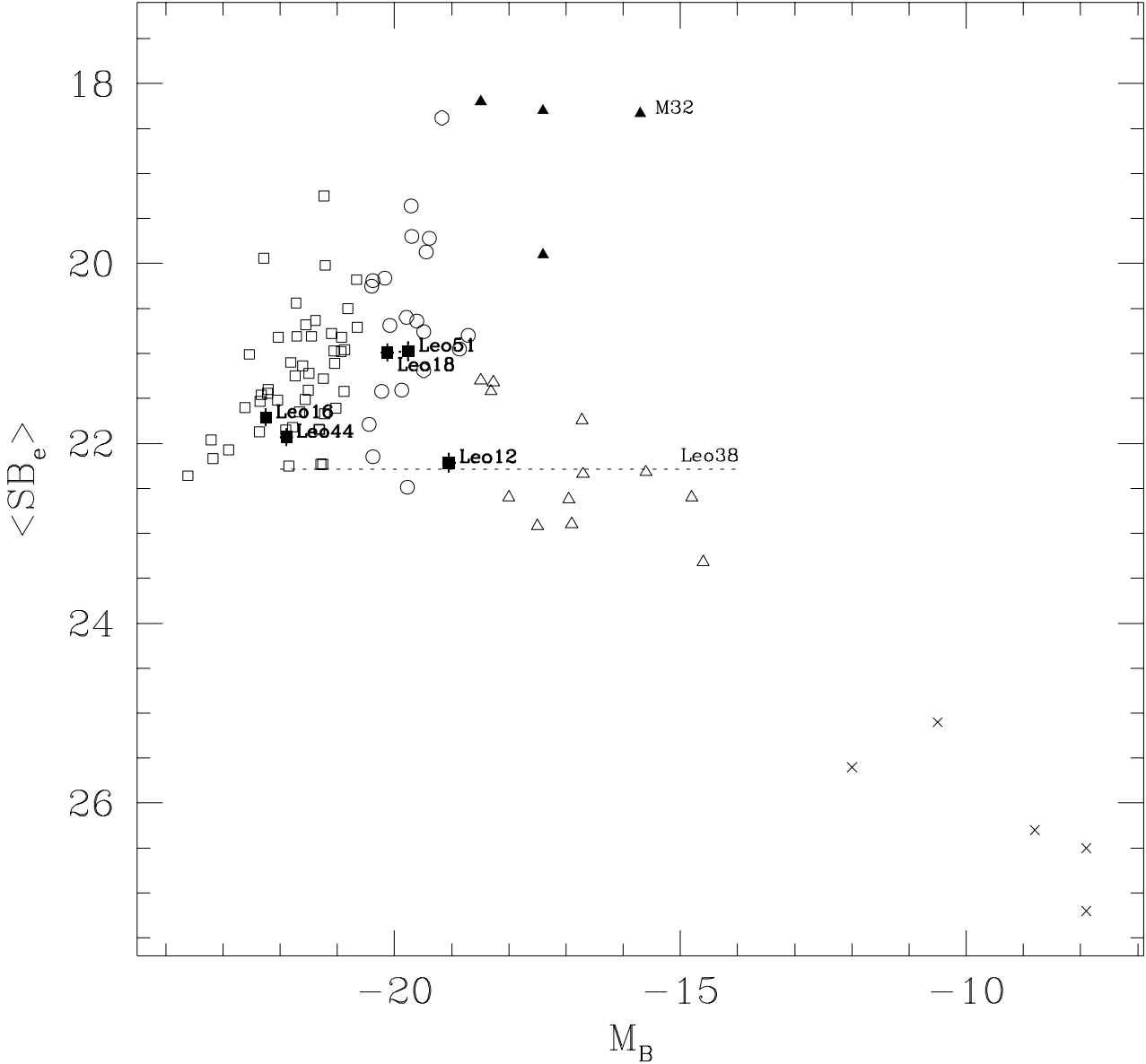,width=18cm}
\caption[]{The location of the {\sf Leo} galaxies (filled squares) in the
surface brightness vs. absolute magnitude plane in comparison to the BBF
sample: giant Es (open squares), intermediate Es (open circles), bright dEs
(open triangles), compact Es like M\,32 (closed triangles) and dSphs
(crosses). The dashed line indicates possible positions of {\sf Leo\,38}. To
calculate $M_B$, $H_o=50$\,km\,s$^{-1}$\,Mpc$^{-1}$ was used}
\label{sbm}
%} \end{center} 
\end{figure*}

The {\sf Leo} galaxies have all non-zero redshifts already indicating
brightnesses too high for dwarf galaxies and placing them well beyond the
Leo group of galaxies. All measured quantities lie within the parameter
range typical for normal elliptical galaxies. In particular, the {\sf Leo}
galaxies obey the well-known Faber-Jackson ($M_B-\sigma$), Fundamental Plane
and Mg$-\sigma$ relation of elliptical galaxies. As an example, we show the
location of the {\sf Leo} galaxies in the surface brightness-absolute
magnitude diagram in Fig.\,\ref{sbm} in comparison to the compilation of
dynamically hot galaxies given by Bender et al. (1993, BBF). \nocite{BBF93}
The mean effective surface brightness within $R_e$ was computed according to
$\langle\mbox{SB}_e\rangle=B+5\lg(R_e)+2.5\lg(2\pi)-10\lg(1+z)$ with $B$
being the total apparent $B$ magnitude. It is evident that the {\sf Leo}
galaxies are not similar to M\,32, but are giant or intermediate ellipticals
in the nomenclature of BBF. {\sf Leo\,12} could also well be classified as a
bright dwarf elliptical. {\sf Leo\,38} has a much too low surface brightness
to be an M32 analogue and, so, even without a spectrum we can safely
conclude that it is either a high luminosity giant elliptical at
$z\approx0.04$ or a dwarf spheroidal at $z\la0.01$.

\begin{figure*}
%\begin{center} \parbox{18cm}{ 
\psfig{figure=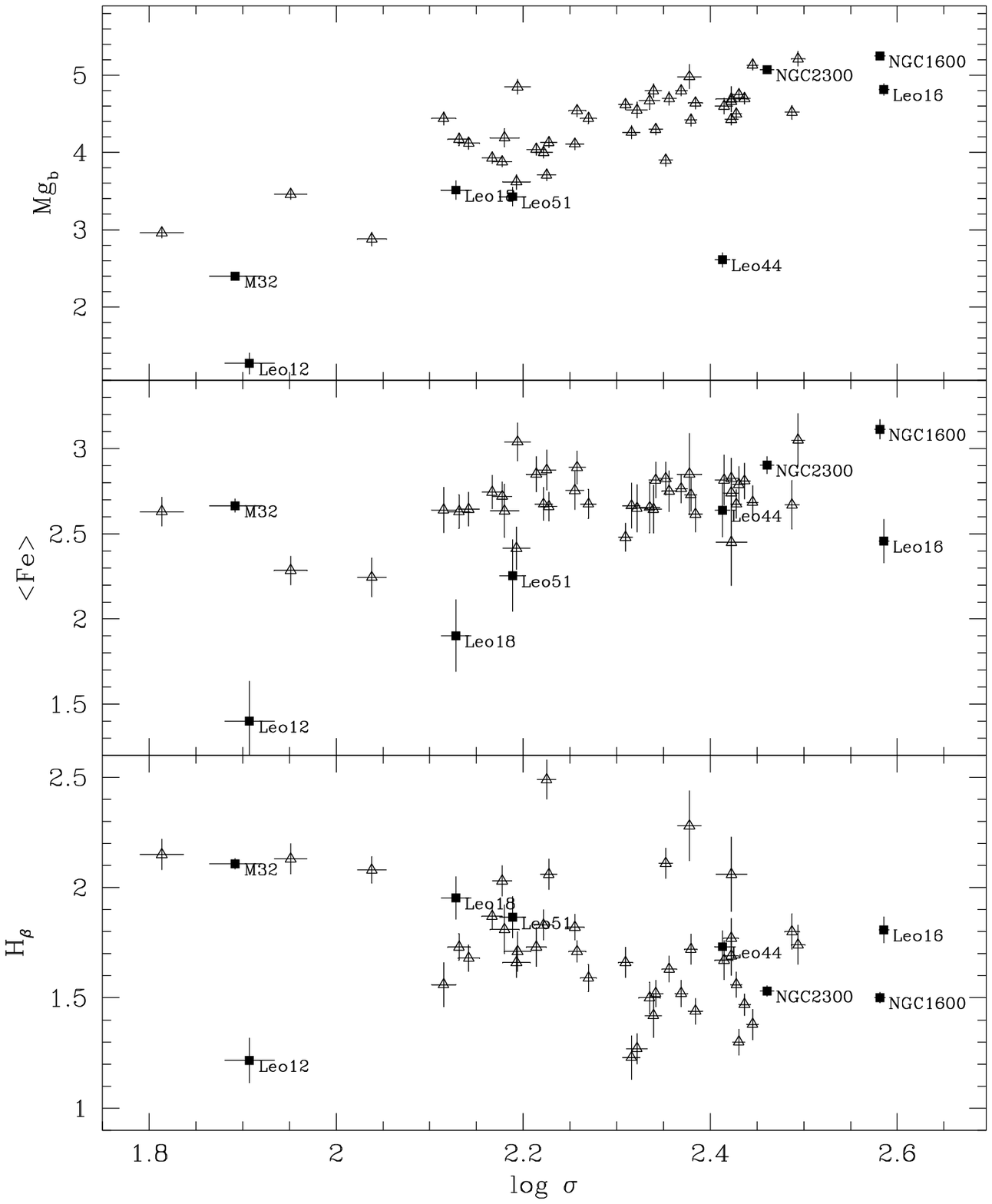,width=18cm}
\caption[]{The measured absorption indices of the {\sf Leo} galaxies (filled
squares) in comparison to the Gonz\'alez sample (open triangles).
$\langle\mbox{Fe}\rangle$ is the arithmetic mean of \fef\ and \fes}
\label{ind}
%} \end{center} 
\end{figure*}

In Fig.\,\ref{ind}, we compare the absorption indices of the {\sf Leo}
galaxies to the sample of elliptical galaxies given by Gon\-z\'a\-lez
\cite*{Gonza93}. Most of the measured values agree well with the
Gon\-z\'a\-lez sample for the same velocity dispersion. The velocity
dispersion of {\sf Leo\,12} is so low that our instrumental resolution is
insufficient to give an accurate value. Its small absorption indices
indicate also the presence of a substantial stellar disk component that
fills \hb\ with emission and produces an [O\,{\sc iii}] emission line.
Therefore, {\sf Leo\,12} might even be a spiral galaxy seen face-on. The
\mgb\ absorption of {\sf Leo\,44} is extraordinarily low which indicates low
age and may be due to a disk component. This is supported by the \hg\
absorption that is just redshifted enough to be visible at the blue end of
the spectrum.

Both, the spectroscopic and the photometric parameters of the observed {\sf
Leo} galaxies demonstrate that they are not M32-analogues. M\,32 is still a
quite unique object with only very few other galaxies having similar
properties and constituting the family of compact ellipticals, however none
of these is as faint as M\,32. If this family is taken as the natural
extension of all elliptical galaxies to lower luminosities the absence of
M32-type objects in Leo adds further evidence that the faint end of the
luminosity function of elliptical galaxies falls off very steeply at
absolute blue magnitudes fainter than --17 \cite{BST88}. Our results also
show that in general only spectroscopy can prove whether a galaxy is of the
type of M32. 
%Therefore, we will try and observe spectroscopically the M32
%candidate galaxies in the other groups given by Ferguson \& Sandage
%\cite*{FS90}, too.

\begin{acknowledgements}
This research was partially supported by the Sonderforschungsbereiche 328
and 375 and DARA grant 50\,OR\,9608\,5.
\end{acknowledgements}

%\bibliography{/usr/users/ziegler/tex/bib/abb,/usr/users/ziegler/tex/bib/all}

\begin{thebibliography}{}

\bibitem[\protect\astroncite{Bender}{1990}]{Bende90a}
Bender, R.: 1990,
\newblock {\rm A\&A} {\rm 229}, 441

\bibitem[\protect\astroncite{Bender et~al.}{1993}]{BBF93}
Bender, R., Burstein, D., and Faber, S.~M.: 1993,
\newblock {\rm ApJ} {\rm 411}, 153

\bibitem[\protect\astroncite{Bender and M{\"o}llenhoff}{1987}]{BM87}
Bender, R. and M{\"o}llenhoff, C.: 1987,
\newblock {\rm A\&A} {\rm 177}, 71

\bibitem[\protect\astroncite{Bender et~al.}{1996}]{BZB96}
Bender, R., Ziegler, B., and Bruzual, G.: 1996,
\newblock {\rm ApJ} {\rm 463}, L51

\bibitem[\protect\astroncite{Binggeli et~al.}{1988}]{BST88}
Binggeli, B., Sandage, A., and Tammann, G.~A.: 1988,
\newblock {\rm ARA\&A} {\rm 26}, 509

\bibitem[\protect\astroncite{Bruzual and Charlot}{1993}]{BC93}
Bruzual, G.~A. and Charlot, S.: 1993,
\newblock {\rm ApJ} {\rm 405}, 538

\bibitem[\protect\astroncite{Christian et~al.}{1985}]{CABBHMS85}
Christian, C.~A., Adams, M., Barnes, J.~V., Butcher, H., Hayes, D.~S., Mould,
  J.~R., and Siegel, M.: 1985,
\newblock {\rm PASP} {\rm 97}, 363

\bibitem[\protect\astroncite{Davidge}{1991}]{David91}
Davidge, T.~J.: 1991,
\newblock {\rm AJ} {\rm 102}, 896

\bibitem[\protect\astroncite{Davies et~al.}{1987}]{DBDFLTW87}
Davies, R.~L., Burstein, D., Dressler, A., Faber, S.~M., Lynden-Bell, D.,
  Terlevich, R.~J., and Wegner, G.: 1987,
\newblock {\rm ApJS} {\rm 64}, 581

\bibitem[\protect\astroncite{Djorgovski}{1992}]{Djorg92}
Djorgovski, S.: 1992,
\newblock in G. Longo et~al. (eds.), {\rm Morphological and Physical
  Classification of Galaxies}, p. 337, Kluwer, Dordrecht

\bibitem[\protect\astroncite{Faber et~al.}{1985}]{FFBG85}
Faber, S.~M., Friel, E.~D., Burstein, D., and Gaskell, C.~M.: 1985,
\newblock {\rm ApJS} {\rm 57}, 711

\bibitem[\protect\astroncite{Ferguson and Sandage}{1990}]{FS90}
Ferguson, H.~C. and Sandage, A.: 1990,
\newblock {\rm AJ} {\rm 100}, 1

\bibitem[\protect\astroncite{Gonz\'alez-Gonz\'alez}{1993}]{Gonza93}
Gonz\'alez-Gonz\'alez, J. d.~J.: 1993,
\newblock Ph.{D}. thesis, {University of California, Santa Cruz}

\bibitem[\protect\astroncite{Horne}{1986}]{Horne86}
Horne, K.: 1986,
\newblock {\rm PASP} {\rm 98}, 609

\bibitem[\protect\astroncite{Kormendy}{1985}]{Korme85}
Kormendy, J.: 1985,
\newblock {\rm ApJ} {\rm 295}, 73

\bibitem[\protect\astroncite{Kormendy}{1987}]{Korme87}
Kormendy, J.: 1987,
\newblock in S.~M. Faber (ed.), {\rm Nearly Normal Galaxies, From the Planck
  Time to the Present}, p. 163, Springer, New York

\bibitem[\protect\astroncite{Kormendy and Bender}{1994}]{KB94}
Kormendy, J. and Bender, R.: 1994,
\newblock in G. Meylan and P. Prugniel (eds.), {\rm Dwarf Galaxies}, {ESO}
  Conf. Proc. No. 49, p. 161

\bibitem[\protect\astroncite{Nieto and Prugniel}{1987}]{NP87}
Nieto, J.-L. and Prugniel, P.: 1987,
\newblock {\rm A\&A} {\rm 186}, 30

\bibitem[\protect\astroncite{Sandage and Binggeli}{1984}]{SB84}
Sandage, A. and Binggeli, B.: 1984,
\newblock {\rm AJ} {\rm 89}, 919

\bibitem[\protect\astroncite{Ziegler and Bender}{1997}]{ZB97}
Ziegler, B.~L. and Bender, R.: 1997,
\newblock {\rm MNRAS}, {\rm 291}, 527

\end{thebibliography}
%\input{m32_bib}

\end{document}